\documentclass[10pt,prd,twocolumn,showpacs,amsmath,amssymb,aps,floats,floatfix,nofootinbib]{revtex4}%
\usepackage{dcolumn}
\usepackage{bm}
\usepackage{amsmath}
\usepackage{amsfonts}
\usepackage{amssymb}
\usepackage{graphicx}
\usepackage{color}
\usepackage{latexsym}
\usepackage{slashed} 
\makeatother

\begin{document}
\title{Pulsar interpretation for the AMS-02 result}
\author{Peng-Fei Yin, Zhao-Huan Yu, Qiang Yuan and Xiao-Jun Bi }
\affiliation{Key Laboratory of Particle Astrophysics,
Institute of High Energy Physics, Chinese Academy of Sciences,
Beijing 100049, China}

\begin{abstract}
The AMS-02 collaboration has just published a high precision
measurement of the cosmic positron fraction $e^+/(e^- + e^+)$,
which rises with energy from $\sim 5$ GeV to $\sim 350$ GeV. The
result indicates the existence of primary electron/positron
sources to account for the positron excess. In this work, we
investigate the possibility that the nearby mature pulsars are the
primary positron sources. By fitting the data we find that the positrons from a single nearby
pulsar, such as Geminga or Monogem, with the spectral index
$\alpha \sim 2$ can interpret the AMS-02 result. We also
investigate the possibility that high energy positrons are
generated by multiple known pulsars in the ATNF catalogue. Such a
scenario can also fit the AMS-02 data well. Future precise measurements
of fine structures in the positron spectrum would be a support to the
pulsar scenario.
\end{abstract}

\maketitle

\section{Introduction}

The Alpha Magnetic Spectrometer (AMS-02) collaboration has just
published its first result about the positron fraction in cosmic
rays (CRs) with extremely high precision \cite{Aguilar:2013qda}. The
positron fraction rises from $\sim 5$ GeV continuously up to $\sim
350$ GeV, while the slope becomes flat above $\sim 100$ GeV. This
result is consistent with the PAMELA result about the positron
fraction \cite{Adriani:2008zr,Adriani:2010ib}. Many studies show
that primary electron/positron sources beyond the conventional
cosmic ray (CR) model are necessary to explain the PAMELA data
(see e.g. \cite{Liu:2011re}). Astrophysical sources, like pulsars
and pulsar wind nebulae (PWNe) \cite{Hooper:2008kg,Yuksel:2008rf,Profumo:2008ms,Malyshev:2009tw,Kawanaka:2009dk,
Blasi:2010de,Kashiyama:2010ui,Kamae:2010ad,Pato:2010im,Serpico:2011wg,Grasso:2009ma,Hall:2008qu,Delahaye:2010ji}, or dark matter (DM) annihilation/decay
have been widely studied as the primary positron sources
(e.g., \cite{review}).

Along the same line several works have appeared to explain the
AMS-02 data \cite{Yuan:2013eja,recent}. DM is still an attractive
interpretation. As shown in Ref.~\cite{Yuan:2013eja}, DM annihilates
into $\tau^+\tau^-$ final states which results in a soft
positron spectrum can account for the AMS-02 data quite well.
Other interpretations that DM annihilation/decay into multiple
$\mu$ or $\tau$ leptons may also be fine to reproduce AMS-02 data.
The DM annihilation scenarios require a large boost factor
and the ``leptophilic'' property of DM particles. However, significant
secondary gamma rays are induced by cascade decay, final state
radiation and inverse Compton scattering. Therefore, it is strongly
constrained by the Fermi-LAT gamma ray observations from the
Galactic center \cite{Hooper:2012sr,Huang:2012yf} or from dwarf
galaxies \cite{Abdo:2010ex,Tsai:2012cs}.

Pulsars are known to be powerful sources in the Galaxy to produce
high energy electrons/positrons with energies of TeV scale or
above \cite{Goldreich:1969sb,Aharonian:1995zz,Atoian:1995ux,Zhang:2001A&A.368}. Primary electrons are extracted from the surface of the
pulsar and are accelerated in the magnetosphere by strong electric
fields (e.g. $10^{12}$ V or higher). Energetic curvature radiation
is emitted in the strong magnetic field, which will result in
electron-positron pairs due to the interactions with magnetic fields
or low energy photons. The high energy gamma-ray photons induced by
those electron-positron pairs from pulsars have been observed
by Fermi-LAT \cite{Abdo:2009ax}.

Unlike the contributions from DM which are assumed to be continuously
distributed in the halo and independent of time, the positron injections
from pulsars are discrete in the Galactic disk and time-dependent.
Since electrons/positrons lose energy quickly via synchrotron radiation
and inverse Compton scattering when propagating in the Galaxy,
the observed electrons/positrons above $\sim 100$ GeV can only come
from a small range within a few kpc. A few nearby mature pulsars may have
very important contributions to the high energy electron/positron
spectrum and may induce significant deviation from the scenarios with
continuous and steady injection.

In this work, we investigate the possibility that nearby mature
pulsars are the sources of the observed high energy positions. We
use a Markov Chain Monte Carlo (MCMC) method to fit the AMS-02
positron fraction data and determine the model parameters
\cite{Liu:2011re}. To determine the properties of the electron
backgrounds, we also include the PAMELA electron data
\cite{Adriani:2011xv} in the fit. We consider the possibilities
that a single nearby pulsar such as Geminga or Monogem is the
source to produce the observed positrons. Through fitting to the
data we get the constraints on the parameters of a single pulsar,
such as the distance, age and total injected $e^\pm$ energy. High
energy $e^\pm$ may also be generated by multiple pulsars rather
than a certain single pulsar. We then discuss the positron
spectrum from a population of pulsars based on the ATNF pulsar
catalogue \cite{Manchester:2004bp}. The multiple pulsars
may produce bump like structures in the positron spectrum.
We discuss the possibility to distinguish the pulsar and
the DM scenarios by a future experiment, such as the Chinese
satellite experiment DArk Matter Particle Explorer (DAMPE), which
is planned to be launched in 2015 \cite{DAMPE}.

 It is worth emphasizing that the number and the energy
distribution of $e^\pm$ pairs injected from the pulsar
magnetosphere are still open questions. Since there may exist a
PWN between the pulsar and the interstellar medium (ISM) (or a
supernova remnant), the spectrum of $e^\pm$ pairs from the pulsar
magnetosphere would be modified by the termination shock and
radiation cooling before they are injected into the ISM
\cite{Blasi:2010de,Serpico:2011wg}. In comparison with the DM
scenario, it is very difficult to obtain a concrete form of the
initial $e^\pm$ spectrum from pulsar models. Therefore, there
would be large uncertainties in the pulsar scenario to explain the
high energy positron excesses.

The paper is organized as follows. In Sec.~II we describe our
treatments for the CR backgrounds and propagation parameters. In
Sec.~III, we discuss the injection $e^\pm$ spectra from the pulsar
and their propagations. In Sec.~IV, we consider the possibility of
a single pulsar as the source of the observed high energy
positrons, and take Geminga and Monogem as benchmark examples. In
Sec.~V, we calculate the positron contributions from multiple
pulsars. Then we discuss the possibility to distinguish the pulsar
scenario from the DM scenario by a future experiment in Sec VI.
Finally Sec.~VII is our conclusion and discussion.

\section{Cosmic ray backgrounds}

The background to explain data includes primary electrons from the
CR sources, and the secondary positrons/electrons
which are generated in the collisions between the CR nuclei
and the ISM. In this work, we use the GALPROP code to calculate the
background \cite{Strong:1998pw,Moskalenko:1997gh}. We employ the
diffusion reacceleration model for CR propagation. The propagation
parameters are adopted by  fitting to the Boron-to-Carbon ratio
and unstable-to-stable Beryllium ratio \cite{mcmc:prop} (see also \cite{Delahaye:2007fr}). The
parameters are $D_0|_{R_0=4\,{\rm GV}}=5.94\times 10^{28}$ cm$^2$
s$^{-1}$, $\delta=0.377$, $v_A=36.4$ km s$^{-1}$ and $z_h=4.04$
kpc. With these propagation parameters, the injection spectrum of
the protons is fitted according to the PAMELA
\cite{Adriani:2011cu} and CREAM \cite{Ahn:2010gv} data. The
fitting parameters of the proton injection spectrum are
$\nu_1=1.80$, $\nu_2=2.36$ and $R_{\rm br}^p=11.7$ GV, where
$\nu_1$ and $\nu_2$ are the spectral indices below and above the
break rigidity $R_{\rm br}^p$ \cite{Yuan:2013eja}.

The secondary positrons and electrons can be calculated according
to the proton spectrum and the propagation model (see e.g. \cite{Delahaye:2008ua}). To involve some
unknown uncertainties, e.g., from the ISM density distribution, the
hadronic interactions and the nuclear enhancement factor from the
heavy elements, we rescale the calculated fluxes of secondary
electrons and positrons with a free factor $c_{e^+}$ in order to
fit the data. The injection spectrum of the primary electrons is parameterized
by a broken power-law with respect to the rigidity (or momentum),
$q(R)\propto(R/R_{\rm br}^e)^{-\gamma_1/\gamma_2}$, with
$\gamma_1$ and $\gamma_2$ the indices below and above the break
rigidity $R_{\rm br}^e$. A further normalization factor $A_e$ is
needed in the fitting procedure.

For energies less than several tens of GeV, the fluxes of CR particles
will be modulated by the solar environment, known as solar modulation.
The force field approximation, with only one parameter $\phi$, is
used to take into account the solar modulation effect
\cite{Gleeson:1968zza}. Note the low energy part ($\lesssim5$ GeV)
of the positron fraction may not be easily explained with the
simple solar modulation model, and more complicated charge-sign
dependent modulation is necessary \cite{DellaTorre:2012zz,
Maccione:2012cu}.

\section{High energy $e^\pm$ pairs from the pulsar}

The energy of $e^\pm$ injected into the ISM is limited by the rotational
energy loss rate of the pulsar. The rotational frequency $\Omega=2 \pi/P$,
with $P$ the pulse period, decreases as $\dot{\Omega}=-a \Omega^n$.
Here $n=\Omega \ddot{\Omega}/\dot{\Omega}^2$ is the breaking index which
can be calculated from the measurements of $\Omega$, $\dot{\Omega}$ and
$\ddot{\Omega}$. The upper limit of the total $e^\pm$ energy is determined
by the pulsar spin-down luminosity $\dot{E}=I\Omega |\dot{\Omega}|=a
\Omega^{n+1}$, where $I=(2/5)M_s R_s^2$ is the moment of inertia of the
pulsar, $M_s$ and $R_s$ are the mass and radius of the pulsar,
respectively. For the magnetic dipole radiation the braking index $n=3$,
and the rotational frequency $\Omega$ is given by (see Ref. \cite{Hooper:2008kg,Profumo:2008ms} and references therein)
\begin{equation}
\Omega(t)=\Omega_0\left(1+\frac{t}{\tau_0}\right)^{-1/2},
\label{frequency}
\end{equation}
where $\Omega_0$ is the initial rotational frequency of the pulsar,
$\tau_0=3c^3 I/ B_s^2 R_s^6 \Omega_0^2$ ($B_s$ is the surface magnetic
field) is a characteristic time scale describing the spin-down luminosity
decays. $\tau_0$ is usually assumed to be $\sim 10^4$ year \cite{Atoian:1995ux}. The spin-down
luminosity of a pulsar is then
\begin{equation}
\dot{E}(t)=\frac{I\Omega_0^2}{2\tau_0}\left(1+\frac{t}{\tau_0}\right)^{-2}.
\label{sdlumi}
\end{equation}
For the pulsar with $t\gg \tau_0$, $\dot{E}$ decreases as $ t^{-2}$.
The age of the pulsar $T$ can be obtained through the rotational energy
loss rate approximately, $T=\Omega/2|\dot{\Omega}|$. The total energy of electrons
and positrons injected from a pulsar is assumed to be proportional to
the rotational energy loss, which is
\begin{equation}
E_{\rm out}=\eta_{e^\pm} \int \dot{E} dt\simeq \eta_{e^\pm} \dot{E}
\frac{T^2}{\tau_0},
\label{etot}
\end{equation}
where $\eta_{e^\pm}$ is the fraction of the rotational energy converted into the energy of
electrons and positrons.

The propagation of high energy $e^\pm$ can be described by the diffusion
equation in the spherically symmetric approximation \cite{Aharonian:1995zz,Atoian:1995ux}
\begin{equation}
\frac{\partial f}{\partial t}=Q(E,t)+ \frac{D(E)}{r^2}
\frac{\partial}{\partial r} r^2 \frac{\partial f}{\partial r} +
\frac{\partial}{\partial E}\left[b(E)f\right],
\label{diffusion1}
\end{equation}
where $f(r,t,E)$ is the time dependent differential density of electrons
and positrons (the differential flux is $c f(r,t,E)/4\pi$), $D(E)\propto\beta D_0 (E/E_0)^\delta $ is the diffusion
coefficient with $D_0$ and $\delta$ the same as the background calculation
(Sec. II). The energy loss rate due to synchrotron and inverse Compton
scattering is adopted as $b(E)=b_0 E^2$ with $b_0=1.4 \times 10^{-16}
~\mathrm{GeV}^{-1}~\mathrm{s}^{-1}$ \cite{Grasso:2009ma}.

For the burst-like source, the source term can be taken as a $\delta$
function $Q(r,t,E) \propto \delta (r-r_0) \delta(t-t_0)$. This is a good
approximation for the pulsar with $T \gg \tau_0$ since most of the
rotational energy is lost during the time scale $\tau_0$. The injection
energy spectrum of the pulsar is parameterized as a power-law function
with an exponential cutoff
\begin{equation}
Q(E,r,t)= Q_0 E^{-\alpha} \exp(-E/E_{\rm cut}) \delta (r-r_0) \delta(t-t_0),
\label{sourcepsr}
\end{equation}
where $\alpha$ is the spectral index, $E_{\rm cut}$ is the cutoff energy,
$Q_0$ is the normalization factor which can be determined by the total
injection energy $E_{\rm out}$. The solution of Eq. (\ref{diffusion1})
for source term Eq. (\ref{sourcepsr}) is \cite{Aharonian:1995zz,Atoian:1995ux}
\begin{eqnarray}
f(d,t_d,E) & = & \frac{Q_0 E^{-\alpha}}{\pi^{3/2} r_{\rm dif}^3}
\left( 1-\frac{E}{E_{\max}} \right)^{\alpha-2} \nonumber \\
& \times & \exp\left( -\frac{E/E_{\rm cut}}{1-E/E_{\rm max}}-
\frac{d^2}{r_{\rm dif}^2} \right),
\label{fluxpsr}
\end{eqnarray}
where $d$ is the distance of the pulsar from the earth and $t_d$ is the
diffusion time into the ISM. Note that $t_d$ may be different from the
actual age of the pulsar $T$ since electrons and positrons may spend some
time in the PWN before their injection in the ISM.
Here we simply assume $t_d\approx T$. $E_{\rm max}\simeq (b_0 t_d)^{-1}$ is
the maximum energy of electrons and positrons surviving from cooling.
For $e^\pm$ with energies larger than $E_{\rm max}$, $f(d,t_d,E)$ is taken
to be $0$. The diffusion distance $r_\mathrm{dif}$ is given by
\begin{equation}
r_{\rm dif}(t_d,E)= 2 \sqrt{D(E)t_d \frac{1-(1-E/E_{\rm max})^{1-\delta}}
{(1-\delta)E/E_{\rm max}} }.
\label{difdpsr}
\end{equation}

Note that for old pulsars with $T \gg 10^5$ yr, the $e^\pm$ injection
energy rate is suppressed by $1/T^2$. In addition, the positrons from
old pulsars should not contribute much to the observed flux at high
energies due to the energy loss in the ISM. The upper limit of the
propagation time of the $e^\pm$ with certain energy is
\begin{equation}
t \lesssim \frac{1}{b_0 E} \sim 2.3 \times 10^5\,\mathrm{yr}
\left( \frac{E}{\mathrm{TeV}} \right)^{-1}.
\label{tlimit}
\end{equation}
For young pulsars with $T \leq O(10^4)$ yr, the positrons may not have
enough time to propagate to the Earth. Moreover, these positrons may
still be trapped in the PWN and not injected in the ISM during such
short time scale. The limit of the propagation time of the $e^\pm$
also suggests an upper limit of the diffusion distance \cite{Kawanaka:2009dk},
\begin{equation}
r \lesssim 2\sqrt{D(E) t} \sim 1\,\mathrm{kpc}
\left( \frac{E}{\mathrm{TeV}} \right)^{-1/3}.
\label{rlimit}
\end{equation}
Thus, the nearby pulsars with ages $T \sim O(10^5)$ yr and distances
$d \lesssim 1$ kpc are thought to be good candidates to interpret the
exotic high energy positrons.

\section{Single pulsar}

\begin{figure*}[!htb]
\centering
\includegraphics[width=0.8\columnwidth]{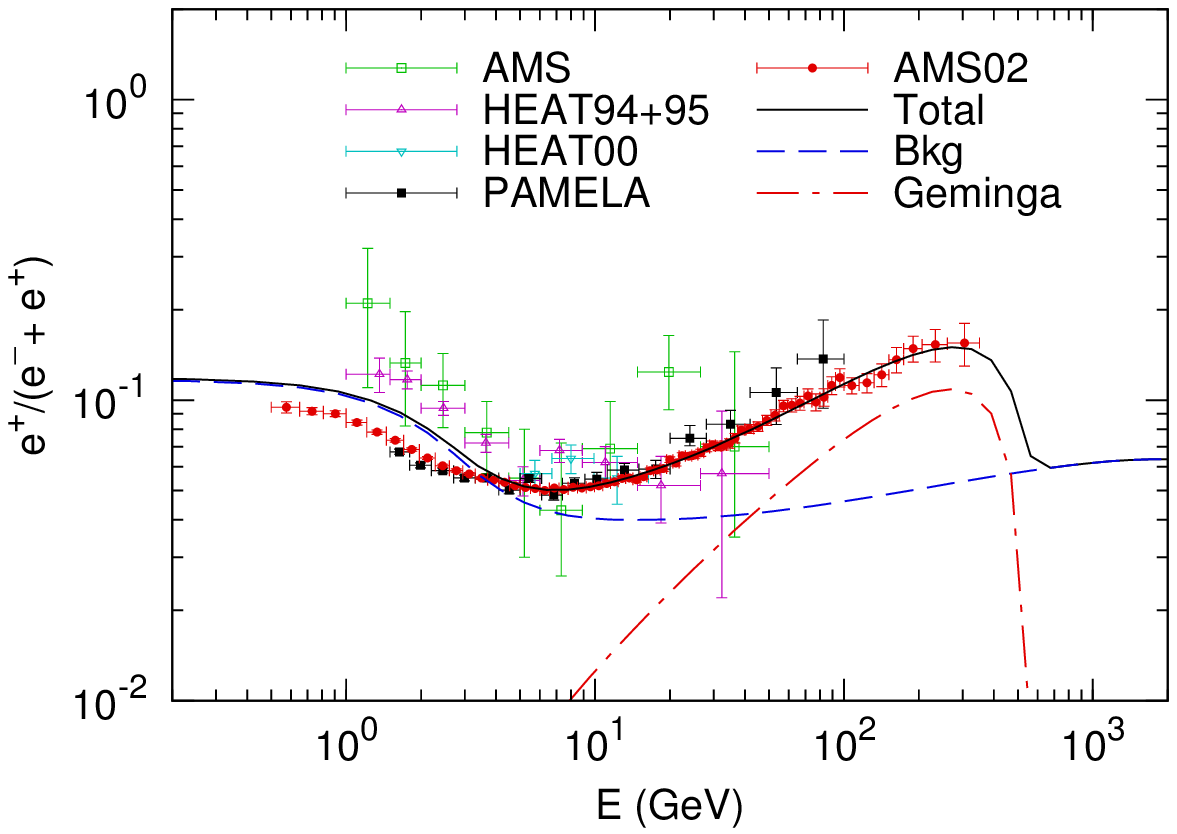}
\includegraphics[width=0.8\columnwidth]{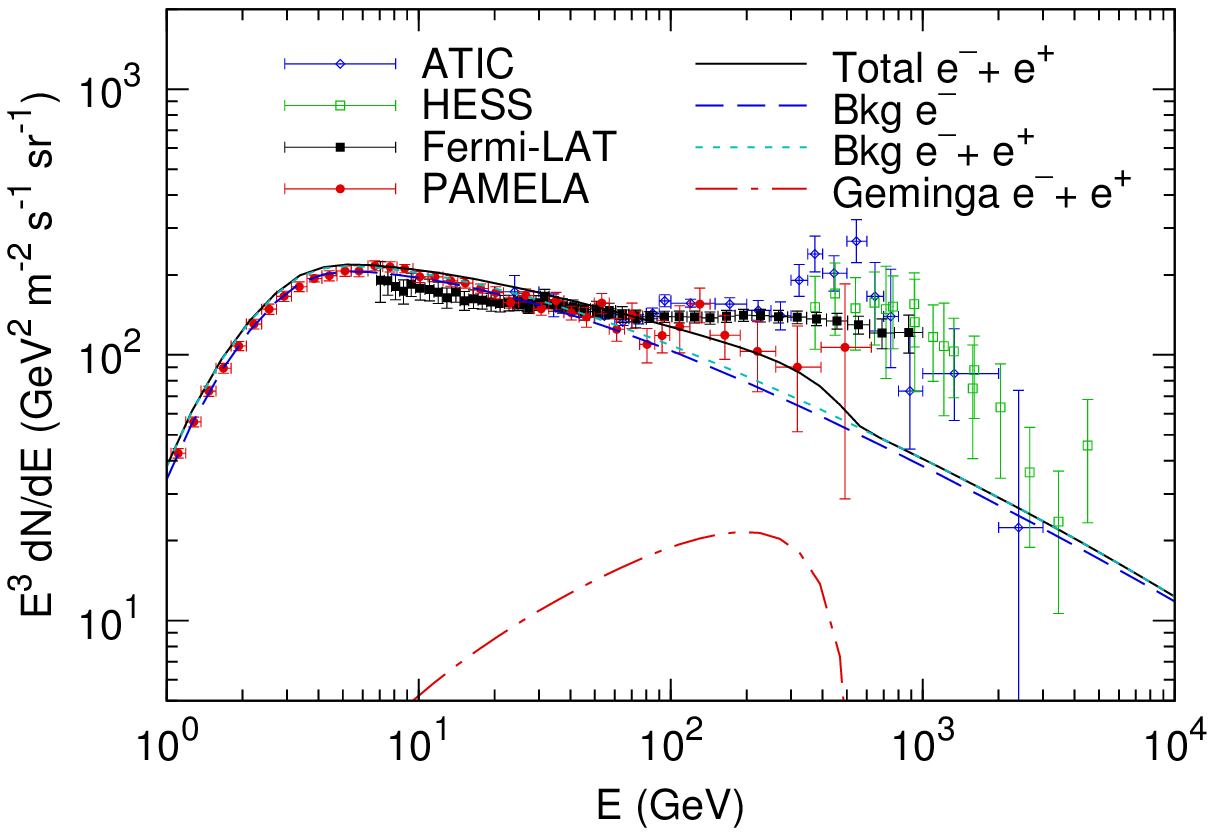} \\
\includegraphics[width=0.8\columnwidth]{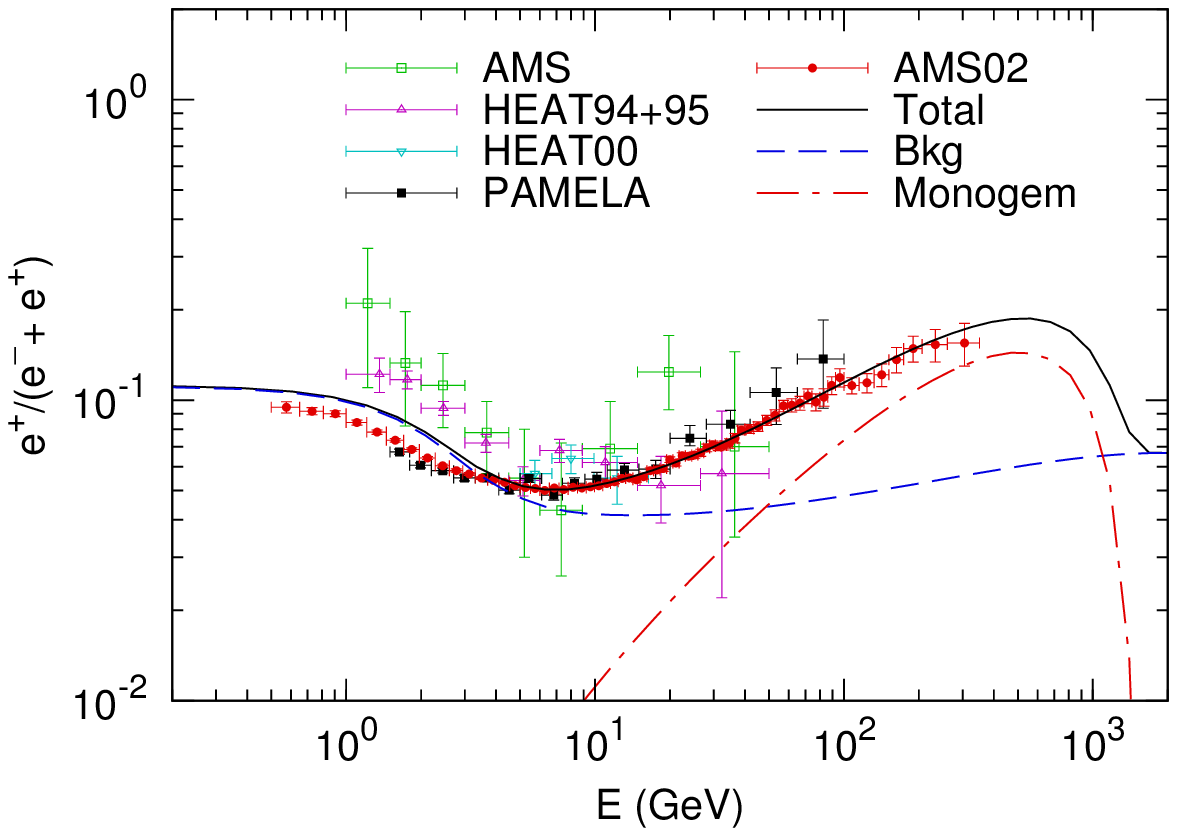}
\includegraphics[width=0.8\columnwidth]{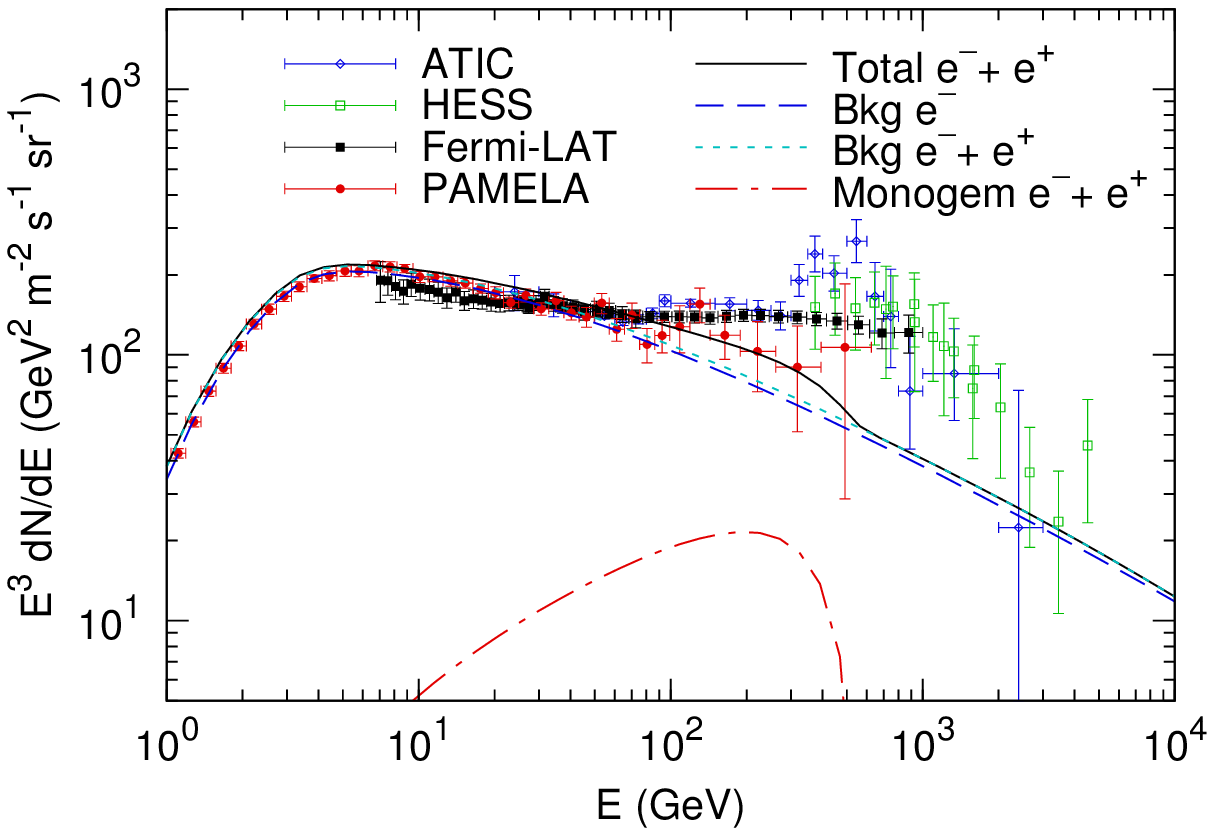}
\caption{The positron fraction (left) and electron flux (right) for the exotic $e^\pm$ from Geminga [J0633+1746] (upper) and Monogem [B0656+14] (lower). Also shown are the positron fraction data from AMS01 \cite{Aguilar:2007yf}, HEAT94+95 \cite{Barwick:1997ig}, HEAT00 \cite{heat00}, PAMELA \cite{Adriani:2008zr} and AMS02 \cite{Aguilar:2013qda}, and electron flux data from PAMELA \cite{Adriani:2011xv}, ATIC \cite{Chang:2008aa}, HESS \cite{Aharonian:2008aa,Aharonian:2009ah} and Fermi-LAT \cite{Ackermann:2010ij}.
\label{psrsin}}
\end{figure*}

In this section, we study the possibility to use a single pulsar
as the high energy positron source to fit the AMS-02 positron
fraction. The background electron spectrum is described by four
parameters, i.e. $A_e$, $\gamma_1$, $\gamma_2$, $R_\mathrm{br}^e$, $c_{e^+}$ and $\phi_e$
which are free parameters in our fit. The positron/electron
spectrum from pulsars are described by five parameters in Eq.~\eqref{fluxpsr},
the distance $d$, the propagation time of $t_d$,
the total injected energy $E_\mathrm{out}$, the cutoff energy $E_\mathrm{cut}$
and the spectral index $\alpha$. For nearby known pulsars, the
distance and age are adopted by the ATNF catalogue data. Note that
the measurements of the pulsar distance still have some
uncertainties. Moreover, the fact that pulsars have velocities of
$\sim O(10^2)$ km s$^{-1}$ means that the current
distance of the pulsar is different from that during the $e^\pm$
injection period (see e.g.~\cite{Profumo:2008ms} and references therein). Here we do not take into account such
uncertainties in our fit.

\begin{figure}[t]
\includegraphics[width=0.9\columnwidth]{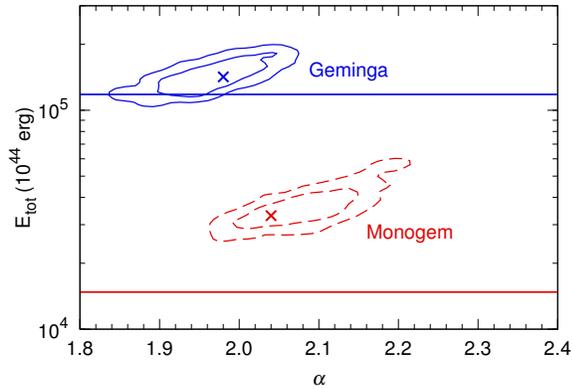}
\caption{$1\sigma$ and $2\sigma$ confidence regions in the plane of $\alpha$ vs. $E_\mathrm{tot}$ for Geminga (blue) and Monogem (red), respectively. Horizontal lines denote the maximum $E_\mathrm{out}$ derived by Eq. (\ref{etot}) with $\tau_0=10^4$ yr and $\eta_{e^\pm}=1$. Also shown are points for best fits.
\label{psrpara}}
\end{figure}

We consider two nearby pulsars Geminga [J0633+1746] with a
distance of $d=0.25$ kpc and age of $T=3.7\times 10^5$ yr, and
Monogem [B0656+14] with $d=0.28$ kpc and $T=1.1 \times 10^5$ yr.
Here we take the age of the pulsar as the diffusion time. We use a
MCMC method to determine $E_\mathrm{cut}$, $\alpha$ and $E_\mathrm{tot}$. The
resulting positron fraction and electron spectrum for the best
fitting parameters for Geminga and Monogem are shown in Fig.~\ref{psrsin}.
The best fitting parameters are given in Table~\ref{bestpara}.
From Fig.~\ref{psrsin} we can see that the pulsar
scenario with these parameters can fit the AMS-02 positron
fraction and the PAMELA electron flux data very well\footnote{Note
that here we have considered the low energy electron data from 1
GeV to 20 GeV in our fit. Since this setting tends to choose a
softer spectrum for the electron background, our fit shows a
tension between PAMELA and Fermi/HESS electron-positron data. More
discussions about this issue can be found in Refs.~\cite{Yuan:2013eja,Yuan:2013eba}.}.
Since Monogem is younger than
Geminga, high energy positrons from Monogem would have smaller
energy loss. Thus, the cutoff of the Monogem spectrum is larger
than that of Geminga. It is possible to observe such cutoff with
more AMS-02 data accumulation.

\begin{table}[th]
\begin{center}%
\begin{tabular}
[c]{c| c c c | c c c c c c}\hline \hline
pulsar & $E_\mathrm{cut}$ & \;\; $\alpha$ \;& $E_\mathrm{tot}\;$ & \;$\log(A_e)$ & $\gamma_1 \;$ & $\; \gamma_2 \;\;$ & $R^e_\mathrm{br}$ & $c_{e^+}$ & $\phi_e$ \\ \hline
Geminga  & 1.0 & 1.98 & 14.2 & -8.93 & 1.74 & 2.75 & 3.61 &  1.53 & 720 \\
Monogem & 0.62 & 2.04 & 3.30 & -8.93 & 1.75 & 2.75 & 3.62 &  1.61 & 735  \\\hline \hline
\end{tabular}
\end{center}
\caption{Parameters of best fit for two nearby pulsars ($E_\mathrm{cut}$, $\alpha$ and $E_\mathrm{tot}$) Geminga (with $d=0.25$ kpc and $T=3.7\times 10^5$ yr), Monogem (with $d=0.28$ kpc and $T=1.1 \times 10^5$ yr), and electron backgrounds ($A_e$, $\gamma_1$, $\gamma_2$, $R^e_\mathrm{br}$, $c_{e^+}$ and $\phi_e$). $R_\mathrm{br}^e$, $E_\mathrm{cut}$ and $E_\mathrm{out}$ are in units of MV, TeV and $10^{48}$ erg, respectively. $A_e$ is normalized at 1 MeV in unit of $\mathrm{cm}^{-3}~\mathrm{s}^{-1}~\mathrm{MeV}^{-1}$.}%
\label{bestpara}
\end{table}

We also show the confidence regions in the plane of $\alpha$ vs.
$E_\mathrm{tot}$ in Fig.~\ref{psrpara}. Compared with the
parameters required to fit the previous PAMELA positron fraction data,
the spectrum of the single pulsar becomes softer. The typical
spectral indices for Geminga to interpret the AMS-02 data is $1.8
\sim 2.1$, while it is $\sim 1.5$ to fit PAMELA positron fraction
data given by Ref. \cite{Hooper:2008kg}. The observed spin-down
luminosity of Geminga is $3.2\times 10^{34}$ erg s$^{-1}$. By
using Eq.~(\ref{etot}) and adopting $\tau_0=10^4$ yr, the total
injection energy to electron and positron pairs is
$1.2\eta_{e^\pm}\times 10^{49}$ erg which is comparable with the
typical values of our best fit.

A similar conclusion can be applied for Monogem.  The typical
injection spectral of Monogem to interpret the AMS-02 data has a
power-law index within $1.9\sim 2.2$ which is softer than that for
the PAMELA positron data. The observed spin-down luminosity of
Monogem is $3.8 \times 10^{34}$ erg s$^{-1}$,
resulting in the total injection energy $1.48\eta_{e^\pm}\times
10^{48}$ erg. The typical value of the $E_\mathrm{tot}$ in our fit is
within $2\sim 6 \times 10^{48}$ erg.

To relax such energy tension, one can change propagation model.
Another possibility is that Geminga or Monogem may not be the only
source to contribute to the observed high energy positrons.

\begin{figure*}[!htb]
\centering
\includegraphics[width=0.9\columnwidth]{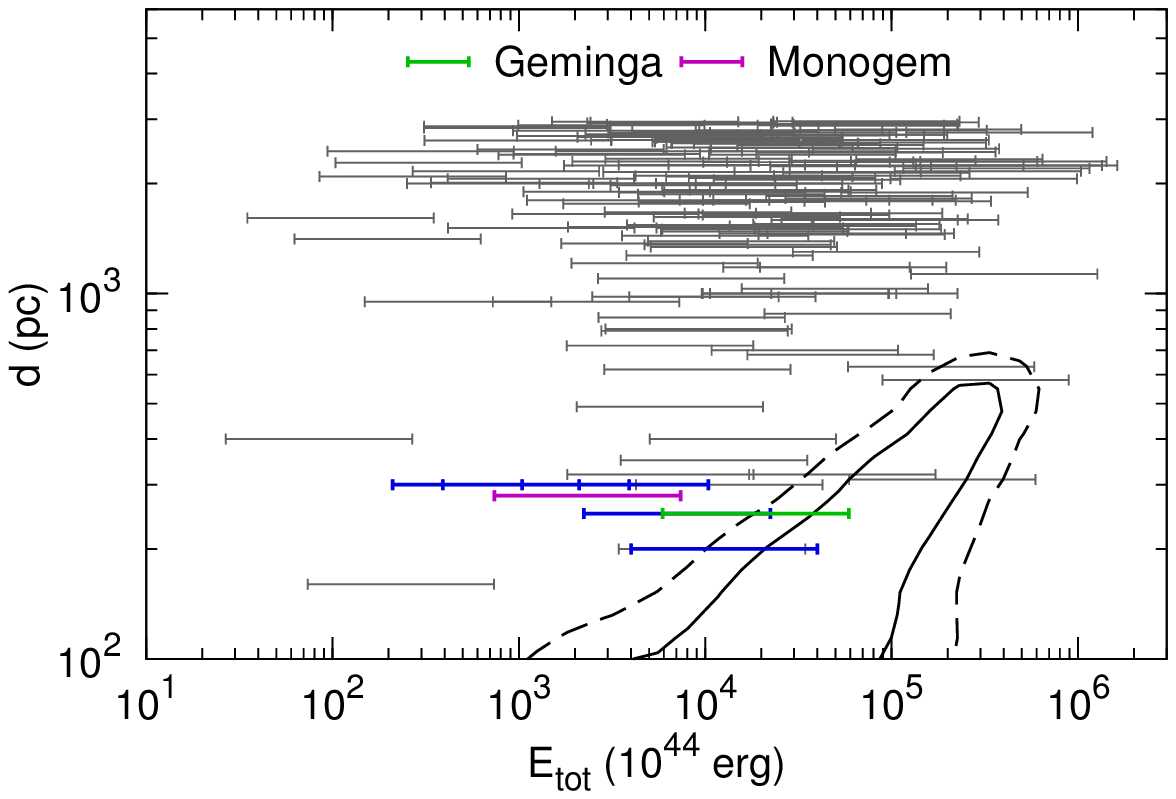}
\includegraphics[width=0.9\columnwidth]{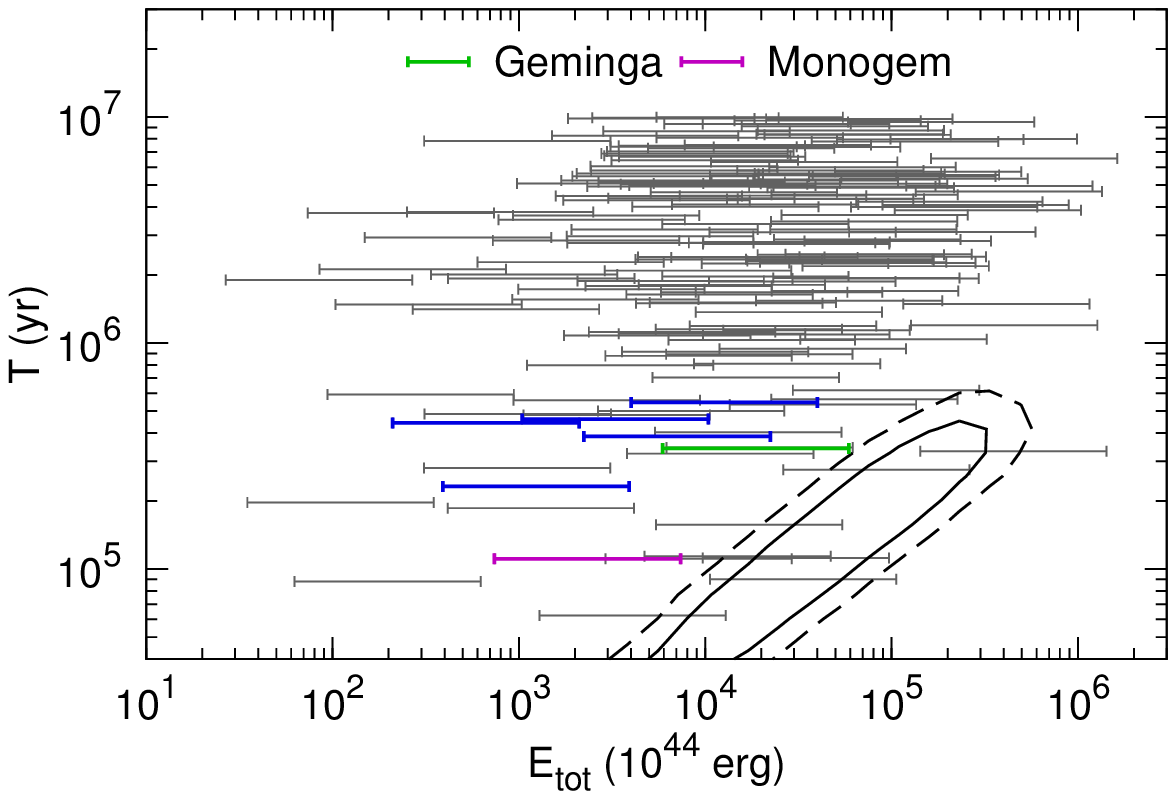}
\includegraphics[width=0.9\columnwidth]{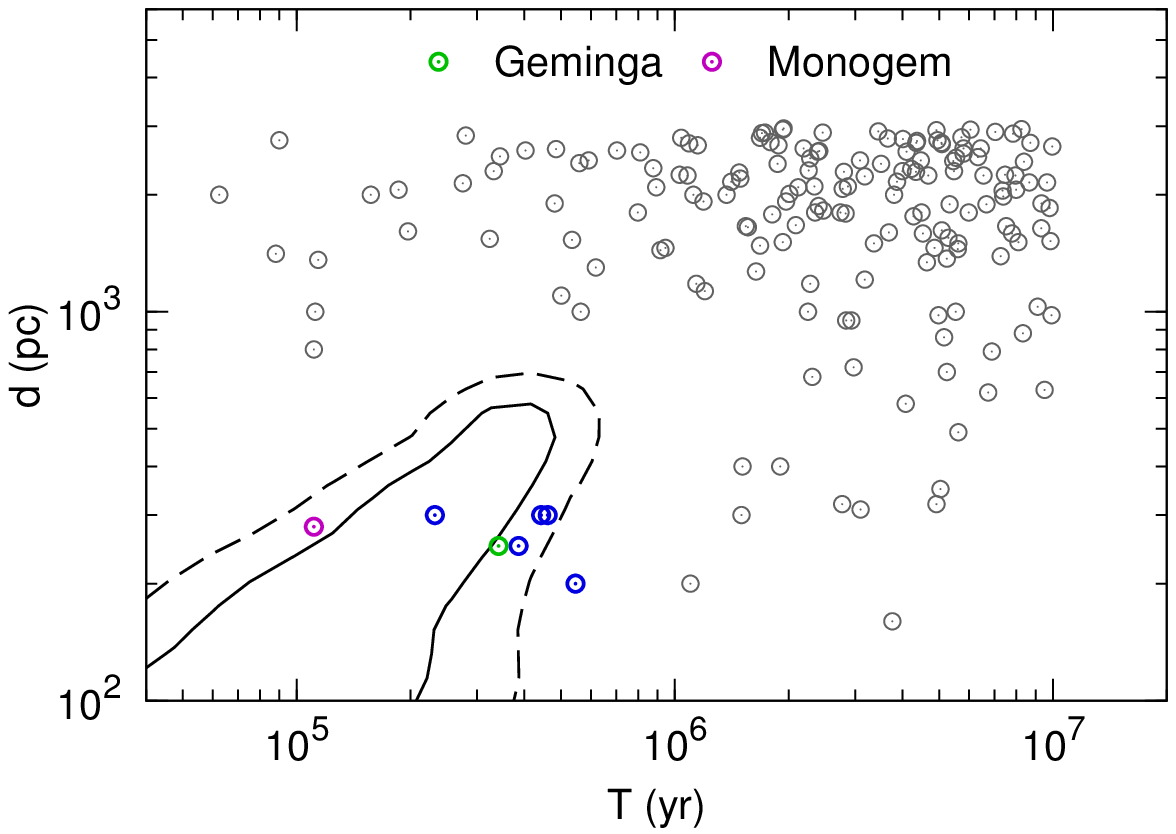}
\caption{$1\sigma$ (solid contour) and $2\sigma$ (dashed contour) confidence regions
in the pulsar parameter space to fit the positron fraction and electron spectrum
for $\alpha=2$ and $E_\mathrm{cut}=1\,\mathrm{TeV}$.
The regions are projected onto the $E_\mathrm{tot}$ vs. $d$ (upper left),
$E_\mathrm{tot}$ vs. $T$ (upper right), and $T$ vs. $d$ (lower) planes.
The circles and bars denote the 177 pulsars in the ATNF catalogue
within $d<3\,\mathrm{kpc}$ and $5\times 10^4<T(\mathrm{yr})<10^7$.
$E_\mathrm{tot}$ for each pulsar is estimated in the range of $5\%\leq\eta_{e^\pm}\leq 50\%$.
7 nearby pulsars with $d<0.5\,\mathrm{kpc}$ and $T<10^6\,\mathrm{yr}$
are marked by colors.
\label{fig:Etot_dist_age}}
\end{figure*}

Then we investigate the correlations between the pulsar distance and age
with fixed spectral index and the cutoff energy following \cite{Pato:2010im}. Applying an MCMC method to fit the
positron fraction and electron spectrum with $\alpha=2$ and
$E_\mathrm{cut}=1\,\mathrm{TeV}$, we derive the $1\sigma$ and
$2\sigma$ confidence regions in the pulsar parameter space, as
shown in Fig.~\ref{fig:Etot_dist_age}.

In Fig.~\ref{fig:Etot_dist_age}, 177 selected pulsars in the ATNF
catalogue with $d<3\,\mathrm{kpc}$ and $5\times
10^4<T(\mathrm{yr})< 10^7$ are also plotted. $E_\mathrm{tot}$ for
each pulsar is estimated from $\dot{E}$ and $T$ by
Eq.~(\ref{etot}) with $\tau_0=10^4\,\mathrm{yr}$ and
$\eta_{e^\pm}$ varying from $5\%$ to $50\%$. From
Fig.~\ref{fig:Etot_dist_age} we note that the 7 nearby pulsars
with $d<0.5\,\mathrm{kpc}$ and $T<10^6\,\mathrm{yr}$ are more
likely to fit the data. We have marked them by colors, with the
color green/magenta corresponding to
Geminga/Monogem.

Fig.~\ref{fig:Etot_dist_age} shows that the favored region in the
pulsar space is rather small. 
There are several pulsars located near the favored
region, especially 7 nearby pulsars marked by colors.
These pulsars could also have sizable contributions to the high
energy $e^\pm$ flux. Therefore, a more reasonable treatment may
include the contributions of all suitable pulsars.

\section{Multiple pulsars}

\begin{figure*}[!htb]
\centering
\includegraphics[width=0.9\columnwidth]{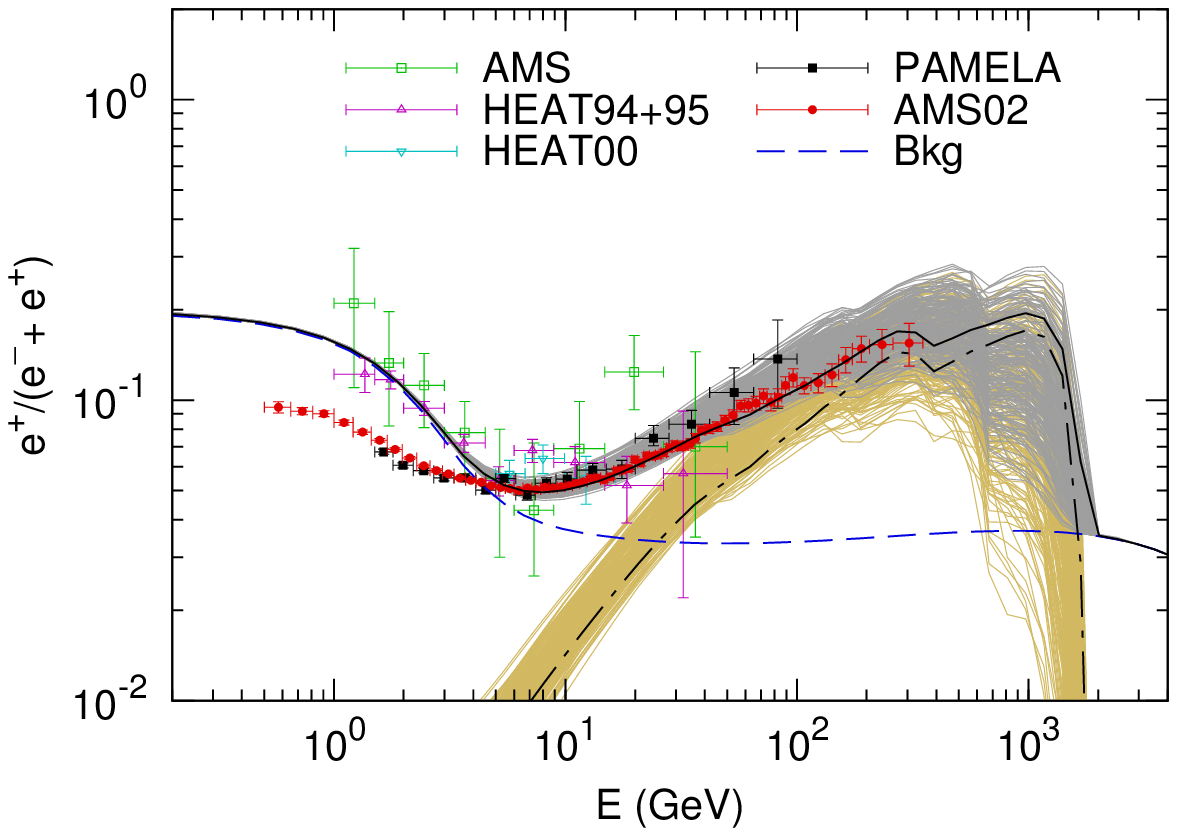}
\includegraphics[width=0.9\columnwidth]{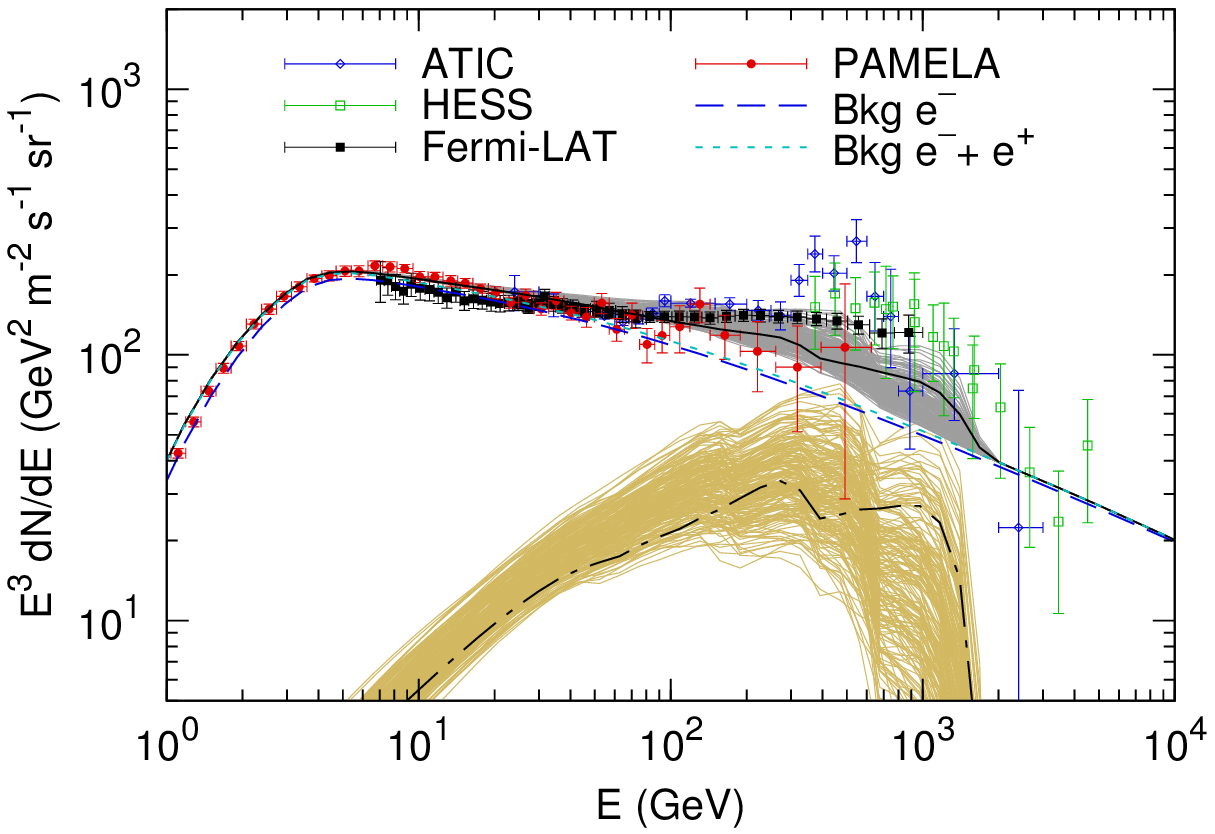}
\caption{The contributions to the positron fraction (left) and electron flux (right)
of all the 177 pulsars in the ATNF catalogue
with $d<3\,\mathrm{kpc}$ and $5\times 10^4<T(\mathrm{yr})<10^7$.
The parameters for each pulsar are randomly assigned in the following ranges:
$700 \leq E_\mathrm{cut}(\mathrm{GeV})\leq 3000$,
$1.5\leq \alpha\leq 2.3$ and $5\%\leq\eta_{e^\pm}\leq 30\%$.
For a particular combination of the parameters, the resulting spectrum
including the contributions of all pulsars is represented by a grey line,
while a golden line shows only the exotic contributions from the pulsars.
A representative choice is shown by black lines
(the solid line for the total result, and the dot-dashed line for the exotic contribution).
\label{fig:random_posi_elec}}
\end{figure*}

It is possible that the flux of high energy electron/positrons are
contributed by many pulsars. Therefore we sum the contribution
from all the 177 mature pulsars in the ATNF catalogue to get the
positron flux following the method in Ref. \cite{Grasso:2009ma}. For each pulsar, we randomly assign the parameters
in the following ranges: $700 \leq
E_\mathrm{cut}(\mathrm{GeV})\leq 3000$, $1.5\leq \alpha\leq 2.3$
and $5\%\leq\eta_{e^\pm}\leq 30\%$. The results are shown in
Fig.~\ref{fig:random_posi_elec}.
Obviously, by summing the contributions of
all pulsar, even low $e^\pm$ pair conversion efficiency
$\eta_{e^\pm}$ could be enough to fit the data.

Since the pulsar parameters vary in large ranges, the resulting total
spectrum of all pulsars also varies in a wide band.
The energy cutoff of each pulsar depends on the minimum of the injection
cutoff $E_{\rm cut}$ and the cooling cutoff $E_{\rm max}$, and is
different from each other. It is further shown that few nearby pulsars
could dominate the total flux. Therefore, for each combination of the
parameters, the sum spectrum tends to have several bumps at high
energies, as shown in Fig.~\ref{fig:random_posi_elec}.

\begin{figure}[!htb]
\centering
\includegraphics[width=0.9\columnwidth]{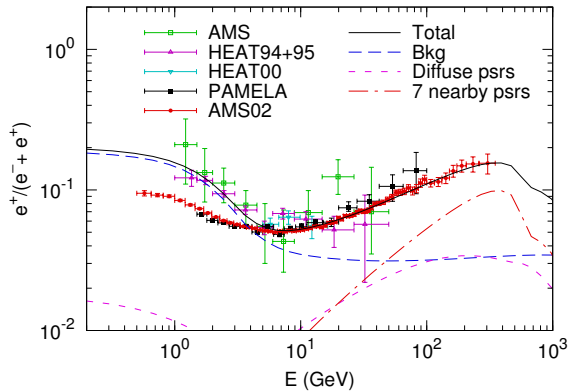}
\caption{Illustration of the positron fraction from both the diffuse
pulsars and the 7 nearby mature pulsars. The parameters of the 7 pulsars
are chosen randomly.
\label{fig:multipsr_posi}}
\end{figure}

Since some pulsars radio beams do not point toward the earth, the ATNF catalogue is incomplete.
There might be a diffuse population of pulsars which are beyond
the observed catalogue. This diffuse component may contribute
as another ``background'' of the electrons/positrons. Similar as done in
\cite{Yuan:2013eja}, we introduce a continuously distributed source
component of the diffuse pulsars, with spatial distribution
\cite{Lorimer:2003qc}
\begin{equation}
Q(R,z)\propto\left(\frac{R}{R_{\odot}}\right)^{2.35}\exp\left[-\frac{5.56
(R-R_{\odot})}{R_{\odot}}\right]\exp\left(-\frac{|z|}{z_s}\right),
\label{source}
\end{equation}
where $R_{\odot}=8.5$ kpc and $z_s\approx 0.2$ kpc. The energy spectrum
of the diffuse pulsars is parameterized by a power-law function with
an exponential cutoff.

Fig.~\ref{fig:multipsr_posi} shows an illustration of the CR positron
fraction from both the diffuse pulsars and the 7 nearby mature pulsars
mentioned above. Here the energy spectrum of the diffuse pulsars is
adopted to be proportional to $E^{-2}\exp(-E/600\,\mathrm{GeV})$.
The parameters $E_\mathrm{cut}$, $\alpha$ and $\eta_{e^\pm}$
of the 7 nearby pulsars are chosen randomly in the ranges described
above. Note that only one combination of the parameters are shown in
in Fig.~\ref{fig:multipsr_posi} as an illustration.

\section{distinguish pulsar from the DM scenario}

\begin{figure*}[!htb]
\centering
\includegraphics[width=0.9\columnwidth]{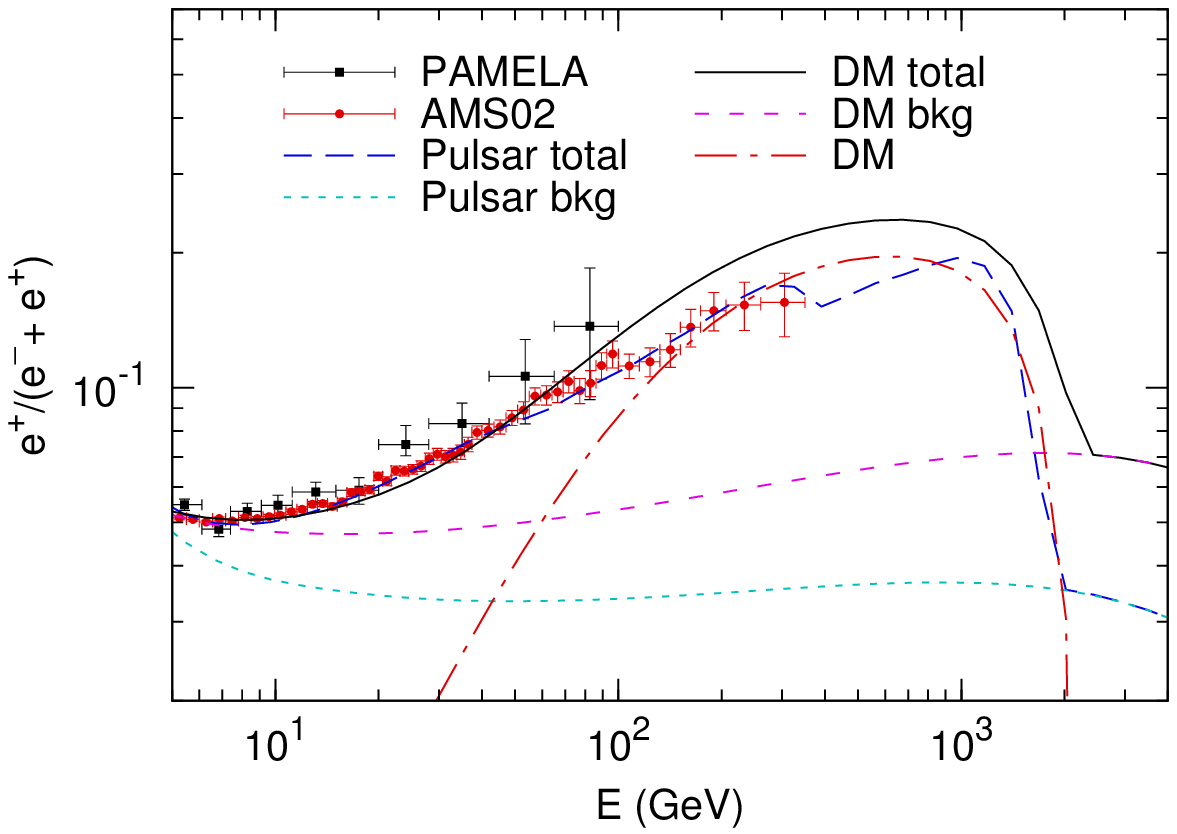}
\includegraphics[width=0.9\columnwidth]{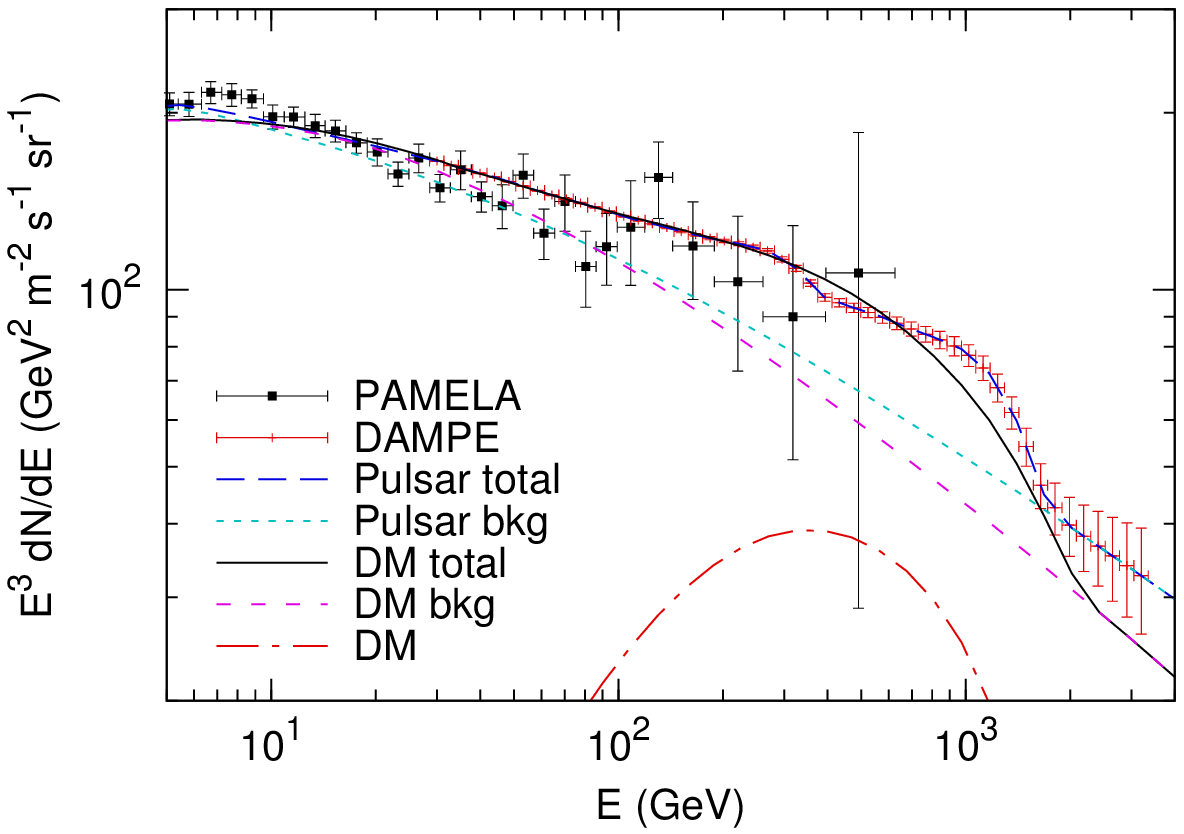}
\caption{The positron fraction (left) and electron
flux (right) for the background together with an exotic component
from multiple pulsars or DM annihilation in $\tau^+\tau^-$ channel.
The ``mock'' DAMPE data are assumed to be contributed by multiple pulsars
and are generated using the pulsar set denoted by black dot-dashed line
in Fig.~\ref{fig:random_posi_elec}.
The DM contribution corresponds to the best fit of an MCMC parameter scan.
\label{fig:mock}}
\end{figure*}

Both the pulsar and the DM scenarios can fit the AMS-02 data well \cite{Yuan:2013eja}. It is a
fundamental problem to distinguish these two
scenarios. As we discussed above if the positron excess is from a
few nearby pulsars, it may have a characteristic spectrum with many
structures. If such fine structures are discovered, it
would be a strong support to the origin of multiple pulsars for the high
energy electrons/positrons.
Here we explore the possibility of distinguishing such two
scenarios by using future electron/positron spectrum observations.
The similar discussions can be found in Refs. \cite{Malyshev:2009tw,Pato:2010im,Hall:2008qu}.

We generate the mock data using the pulsar set denoted by
black dot-dashed line in Fig.~\ref{fig:random_posi_elec} as an
example, and consider the observation capability of DAMPE.
The mock data are produced following the method in Ref.~\cite{Pato:2010im}.
The number of particles detected in a certain energy bin with a gaussian energy resolution is given by
\begin{equation}
N(E)= \Delta t ~\delta E ~A \int d E' \phi (E') \frac{1}{\sqrt{2 \pi \sigma^2}}
\exp{\left[-\frac{(E'-E)^2}{2\sigma^2}\right]},
\end{equation}
where $\phi $ is the differential flux, $\delta E$ is the width of
the energy bin, $\Delta t$ is the observation time, $A$ is the
geometrical factor of the detector and $\sigma$ is determined by
$\sigma= \Delta E/2$. $\Delta E$ is adopted to be the form of
$\Delta E/E = a/\sqrt{E/\mathrm{GeV}}\oplus b$, which is
normalized to $1.5\%$ and $10\%$ at 1 TeV and 1 GeV for DAMPE,
respectively \cite{DAMPE}. The geometrical factor and performing
time of the detector are taken to be
$0.5~\mathrm{m}^2~\mathrm{sr}$ and 5 yr,
respectively~\cite{DAMPE}. The relative statistical uncertainty
can be estimated by $\sqrt{N_e}/N_e$. The systematic uncertainty
is assumed to be mainly determined by the capability of
distinguishing electrons/positrons and other CR particles
\cite{Pato:2010im}. Here the $e/p$ separation is taken to be
$5\times 10^5$ corresponding to the detector thickness of $32$
radiation lengths~\cite{DAMPE}. The relative systematic
uncertainty is estimated by $(N_p/5\times 10^5)/N_e$.

We use the MCMC method to explore the possible DM parameter space to
fit the mock electron/positron flux data for DAMPE,
the positron fraction data from AMS-02 and the electron flux data from PAMELA.
As above, the proton injection spectrum is fixed,
while the parameters of the primary electron injection spectrum are free.
Therefore, the background corresponding to the fit for the DM
is usually different with that for multiple pulsars.
The DM annihilation final states are taken to be $\tau^+\tau^-$,
which induce a soft positron spectrum favored by the AMS-02 result.
From the results shown in Fig.~\ref{fig:mock},
we find that the behavior of the electron/postron spectrum from the DM source is
mainly determined by the mock DAMPE data below $\sim 300$ GeV due to very small uncertainties, and it
cannot reproduce the fine structures above 300 GeV at the data.
If the differences between the electron/positron spectra from the DM and the pulsar origins
are significant as the examples shown here,
DAMPE would have the capability to discriminate these two scenarios.

\section{Discussions and conclusions}

In this work, we investigate the pulsar origin of the positron fraction
measured by AMS-02 recently. We first consider the case that the high
energy positrons are produced by a single pulsar, such as Geminga or
Monogem. We find the AMS-02 data can be well fitted in this case with a
soft power-law index of $\alpha \sim 2$. Such a soft spectrum requires
a large injection energy from the pulsar, which is comparable to the
total energy loss of the pulsar derived from Eq.~(\ref{etot}).
Considering the uncertainties from CR propagation parameters and the
pulsar models, such a tension may be relaxed. We then consider the case
that the positrons are from multiple pulsars in the ATNF catalogue.
We find such scenario can also fit the AMS-02 data very well.

It is shown that pulsars can be a natural explanation of the AMS-02
positron excess. Compared with the DM scenario, pulsar scenario may
have some distinct features to be distinguished from the DM models.
It is very possible that there might be fine structures on the
electron/positron spectrum in the pulsar scenario, because the
parameters of pulsars might differ from one to another
\cite{Malyshev:2009tw}. Furthermore, since one or several nearby
pulsar(s) may dominate the flux of high energy positrons,
it may induce a remarkable anisotropy \cite{Hooper:2008kg,Pato:2010im}. Both the fine structures and
the anisotropy could be tested with future observations.

\begin{acknowledgments}
This work is supported by the Natural Science Foundation of China under the Grant Nos. 11075169,
11105155, 11105157 and by 973 Program under Grant No. 2013CB837000.
\end{acknowledgments}

\;

\emph{Note added}: While this paper was in preparation, two similar
papers on the pulsar interpretation for the AMS-02 result
appeared \cite{Linden:2013mqa,Cholis:2013psa}.  Our results for fitting the AMS-02 data are
consistent with theirs.

\setcounter{equation}{0}
\renewcommand{\theequation}{\arabic{section}.\arabic{equation}}%

\end{document}